\documentclass[12pt,a4paper]{article}
\usepackage[T1]{fontenc}
\usepackage[latin1]{inputenc}

\usepackage{epsfig}
\newcommand{\lsim}{\,{\buildrel < \over {_\sim}}\,}
\newcommand{\gsim}{\,{\buildrel > \over {_\sim}}\,}

\newcommand{\ptD}{p_T^{\scriptscriptstyle D}}

\begin{document}

\begin{titlepage}
\begin{flushright}
HIP-2004-06/TH\\
LBNL-54671\\
hep-ph/0403098
\end{flushright}
\vfill
\begin{centering}

{\bf $D$ MESON ENHANCEMENT IN $pp$ COLLISIONS AT THE LHC
DUE TO NONLINEAR GLUON EVOLUTION}

\vspace{0.5cm}
A. Dainese\renewcommand{\thefootnote}{\arabic{footnote}}\setcounter{footnote}{0}\footnote{andrea.dainese@pd.infn.it}, 
\\
\vspace{0.15cm}

{\em \small University of Padova and INFN, via Marzolo 8, 35131 Padova, Italy}
\vspace{0.3cm}

R. Vogt\footnote{vogt@lbl.gov}, \\
\vspace{0.15cm}

{\em \small 
Lawrence Berkeley National Laboratory, Berkeley, CA 94720, USA \\ 
Physics Department, University of California, Davis, CA 95616,
USA}
\vspace{0.3cm}

M. Bondila\footnote{mariana.bondila@phys.jyu.fi},
K.J. Eskola\footnote{kari.eskola@phys.jyu.fi}, and
V.J. Kolhinen\footnote{vesa.kolhinen@phys.jyu.fi} \\
\vspace{0.15cm}

{\em \small Department of Physics,
P.O. Box 35, FIN-40014 University of Jyv\"askyl\"a, Finland \\
Helsinki Institute of Physics,
P.O. Box 64, FIN-00014 University of Helsinki, Finland}

\vspace{1cm} 
{\bf Abstract} \\ 
\end{centering}

When nonlinear effects on the gluon evolution are included with constraints
from HERA, the gluon distribution in the free proton is enhanced at low 
momentum fractions, $x \, \lsim \, 
0.01$, and low scales, $Q^2\lsim 10$ GeV$^2$, 
relative to standard, DGLAP-evolved, gluon distributions.  Consequently, such
gluon distributions can enhance charm production in $pp$ collisions at center
of mass energy 14 TeV by up to a factor of five at midrapidity,
$y\sim0$, and transverse momentum
$p_T\rightarrow0$ in the most optimistic case.
We show that most of this enhancement survives hadronization into $D$ mesons. 
Assuming the same 
enhancement at leading and next-to-leading order, we show that
the $D$ enhancement may be measured by $D^0$ reconstruction in the $K^-\pi^+$ 
decay channel with the ALICE detector.

\vspace{0.3cm}\noindent

\vfill
\end{titlepage}

\setcounter{footnote}{0}

\section{Introduction}

The parton distribution functions, PDFs, of the free proton are
determined through global fits obtained using the leading-order, LO,
next-to-leading order, NLO, or even next-to-next-to-leading order, NNLO,
formulation of the Dokshitzer, Gribov, Lipatov, Altarelli and Parisi, DGLAP, 
scale evolution equations~\cite{DGLAP}. 
In particular, the HERA data on the proton structure function $F_2(x,Q^2)$
\cite{HERA} as a function of Bjorken-$x$ and
squared momentum transfer $Q^2$, and, especially, the $Q^2$ slope,
$\partial F_2(x,Q^2)/\partial \ln Q^2$, in the
small-$x$, $3\times 10^{-5}\, \lsim x \lsim \,5\times 10^{-3}$,  
and small-$Q^2$ region, $1.5 \, \lsim Q^2 \lsim 10$~GeV$^2$,
set rather stringent constraints on the small-$x$ gluon
distributions. The agreement of the global fits with the measured
$F_2(x,Q^2)$ is, in general, very good but certain problems arise.  
When the small-$x$ and small-$Q^2$ region is included in the DGLAP fits, 
they are not as good as the excellent ones obtained at larger values of $x$
and $Q^2$~\cite{MRS03}. In addition, some NLO gluon distributions
\cite{MRST2001} become negative at small $x$ for $Q^2$ on the order of a few
GeV$^2$.

The kernels of the DGLAP equations only describe splitting of one parton into
two or more so that the resulting equations are linear in the PDFs.  This
ignores the fact that, at low $Q^2$, the small-$x$ gluon density may increase
to the point where gluon fusion becomes significant.  These fusions generate
nonlinearities in the evolution equations.  The first nonlinear corrections, 
the GLRMQ terms, were derived by Gribov, Levin and Ryskin and also
by Mueller and Qiu~\cite{GLRMQ}. Eventually, at even smaller $x$ and $Q^2$,
nonlinearities are expected to dominate the evolution to all orders.  This
fully nonlinear region, where both the linear DGLAP evolution and the 
GLRMQ-corrected  DGLAP evolution are inapplicable, is the gluon saturation 
region, see e.g. Ref.~\cite{EHKQSinHPC}.

Outside the saturation region, incorporating the nonlinearities may improve 
the global fits when the small-$x$ and $Q^2$ regions are included. 
Recent work in Ref.~\cite{EHKQS}, where the LO DGLAP evolution equations 
were supplemented by the GLRMQ terms, showed that the nonlinearly-evolved PDFs
reproduce the HERA $F_2$ measurements at $x \, \gsim \, 3\times 10^{-5}$ and
$Q^2\, \gsim \, 1.5\,{\rm GeV}^2$~\cite{HERA} equally well or even better
than the conventional LO PDFs such as CTEQ6L~\cite{CTEQ6}. The
nonlinearly-evolved gluon distributions at $Q^2\, \lsim \, 10$ GeV$^2$ and
$x\, \lsim \, 0.01$, however, were clearly enhanced relative to CTEQ6L and
CTEQ61L~\cite{cteq61}.  As shown in Fig.~1 of Ref.~\cite{EKV2}, the enhancement
arises because the nonlinear evolution is slower than DGLAP alone.
At higher $x$ and $Q^2$ the nonlinear and linear evolution of the
gluon distributions should become very similar to fit the same data.  
An enhancement can also be expected at NLO. 
However, since the NLO small-$x$ gluon distributions are typically reduced
relative to LO, at NLO the enhancement may be smaller than at LO~\cite{MRS03}.

Since the same HERA data can be reproduced by linear evolution starting from
a relatively flat gluon distribution and by nonlinear evolution with
clearly enhanced small-$x$ gluons, other observables are necessary
to probe the effects of the nonlinearities. In Ref.~\cite{EKV2},
charm production in $pp$ collisions at the LHC was suggested as a
promising candidate process. Due to gluon dominance of charm production
and the small values of $x$ and $Q^2$ probed, $x \approx 2 \times 10^{-4}$ and
$Q^2 \approx 1.69 - 6$ GeV$^2$ at midrapidity and 
transverse momentum\footnote{Here we use $p_T$ for the transverse 
momentum of the charm quark and $\ptD$ for the transverse momentum of the 
$D$ meson.}
$p_T \approx 0$, charm production at
the LHC is sensitive to the gluon enhancement. The resulting charm
enhancement was quantified in Ref.~\cite{EKV2} by the LO ratios
of the differential cross sections computed with the nonlinearly-evolved 
EHKQS PDFs~\cite{EHKQS}, obtained from DGLAP+GLRMQ evolution,
relative to the DGLAP-evolved CTEQ61L PDFs.

The enhancement of the nonlinearly-evolved gluons increases as $x$ and $Q^2$ 
decrease. Consequently, the charm enhancement increases with center of mass
energy, $\sqrt{s}$.  Thus the maximum enhancement at the LHC will be at 
$\sqrt s = 14$~TeV and small charm quark transverse momentum. 
The sensitivity of the charm
enhancement to the value of the charm quark mass, $m_c$, 
as well as to the choice of the factorization, $Q_F^2$,
and renormalization, $Q_R^2$, scales was studied in Ref.~\cite{EKV2}
assuming 
$Q^2 = Q_F^2 = Q_R^2 \propto m_T^2$, the charm transverse mass squared, 
$m_T^2 = p_T^2 + m_c^2$. For the most significant charm enhancement, $m_c$ and
$Q^2/m_T^2$ should both be small.  A comparison of the
NLO total cross sections with low energy data shows that the data prefer such
small $m_c$ and $Q^2$
combinations~\cite{HPC,rvww02}.  The
smallest scales and thus the  largest enhancement are obtained with
$m_c=1.3$~GeV and $Q^2=m_T^2$. 
In this case, the ratio of the inclusive 
differential cross section, $d^3\sigma/dp_T dy dy_2$, computed with
EHKQS set 1 relative to CTEQ61L is greater than 5 for rapidities
$|y, y_2|\, \lsim \, 2$ where $y$ and $y_2$ are the $c$ and $\overline c$
rapidities, respectively.

In Ref.~\cite{EKV2}, the enhancement was described only for charm production.
Neither its subsequent hadronization to $D$ mesons nor its decay and detection
were considered.  
In this paper, we address these issues to determine whether the
charm enhancement survives hadronization and $D$ decay. At the LHC, the ALICE
detector~\cite{aliceTP} 
is perhaps in the best position for measuring such an enhancement 
since it is capable of reconstructing $D^0$ hadronic decays down to very low
transverse momentum.

We first consider how much of the LO charm enhancement survives in the final
state $D$ meson distributions. 
Charm quarks are hadronized using the PYTHIA string
fragmentation model~\cite{pythia}.  We show that, for the most optimistic case
with a factor of five charm enhancement for $p_T \rightarrow 0$, 
the $D$ enhancement is a factor of three for
$\ptD \rightarrow 0$.

Since the ALICE detector allows direct measurement of the $D$ meson $p_T$
distribution through $D^0$ reconstruction in the $K^- \pi^+$ decay
channel, we then determine whether or not the surviving $D$ enhancement can be
detected above the expected
experimental statistical and systematic uncertainties.
To determine realistic statistical uncertainties, we calculate the NLO cross
section in the way most compatible with our LO enhancement, as described below.
Then, using the error analysis
developed by one of us (A.D.) in Ref.~\cite{thesisAD}, we demonstrate that
detection of the enhancement is possible.

Finally, we consider whether NLO charm cross sections,  
calculated with linearly-evolved PDFs and different
combinations of $m_c$, $Q_F^2$ and $Q_R^2$,
can mimic the charm enhancement.  Our
results show that this is unlikely.

This paper is organized as follows.  In Section 2, we describe our charm
calculations and define how the NLO cross section most compatible with the
LO enhancement is computed.  Hadronization and reconstruction of $D^0$ mesons 
are considered in Sections 3 and 4, respectively, along with a discussion of 
the  experimental uncertainties.  We then generate ``data'' based on the 
enhanced cross sections and the experimental uncertainties.  These data are 
then compared to compatible NLO calculations to learn whether the enhancement 
is measurable for a unique set of parameters in Section 5.  We conclude in 
Section 6.

\section{Charm enhancement from nonlinear PDF evolution}

According to collinear factorization, the inclusive differential charm
hadroproduction cross sections at high energies can be written as
\begin{eqnarray}
 d\sigma_{pp\rightarrow c \overline c X}(\sqrt s,m_c,Q_R^2,Q_F^2) 
&=&
   \sum_{i,j=q,\overline q,g} 
   f_i(x_1,Q_F^2)\otimes f_j(x_2,Q_F^2)   \nonumber \\
&& \otimes \, d\hat \sigma_{ij\rightarrow c \overline c \{k\}}
(\alpha_s(Q_R^2),Q_F^2,m_c,x_1,x_2),
\label{sigcc}
\end{eqnarray}
where $d\hat \sigma_{ij\rightarrow c \overline c\{k\} }$ is the perturbative
partonic hard part, calculable as a power series in the strong
coupling $\alpha_s(Q_R^2)$.
The proton PDFs for each parton $i(j)$ at fractional momentum $x_1 (x_2)$
and factorization scale $Q_F^2$
are denoted by $f_i(x,Q_F^2)$. At LO,
where $d\hat\sigma \propto \alpha_s^2(Q_R^2)$, only the subprocesses 
$gg\rightarrow c\overline c$ and $q\overline q\rightarrow c\overline c$ 
are allowed~\cite{COMBRIDGE} so that $\{k\}=0$.  At NLO, 
where $d\hat \sigma \propto \alpha_s^3(Q_R^2)$, 
subprocesses where $\{k\}\ne0$, e.g.\
$gg\rightarrow c\overline c g$ and $gq\rightarrow c\overline c q$ contribute.
The $g q$ channel, new at NLO, only contributes 
a few percent of the total cross section. 
 
The charm production enhancement studied here and in
Ref.~\cite{EKV2} results from the nonlinearly-evolved EHKQS 
PDFs\footnote{These PDFs are available at www.urhic.phys.jyu.fi.} where the
gluon distribution is enhanced 
for $x \, \lsim \, 0.01$ at the few-GeV scales. The EHKQS
PDFs were constructed in Ref.~\cite{EHKQS} using CTEQ5L
\cite{CTEQ5} and CTEQ6L as baselines with the HERA data
\cite{HERA} as constraints.  The EHKQS sets have initial scale 
$Q_0^2 = 1.4$~GeV$^2$ and a
four-flavor $\Lambda_{\rm QCD}$ value of $\Lambda_{\rm QCD}^{(4)} =
0.192$~GeV.  Following Ref.~\cite{EKV2}, we quantify
the charm enhancement against charm production computed with the
CTEQ61L LO PDFs where the data were fit with the one-loop $\alpha_s$.
The CTEQ61L set takes $Q_0^2 = 1.69$~GeV$^2$ and
$\Lambda_{\rm QCD}^{(4)} = 0.215$~GeV.  For consistency, we
calculate $\alpha_s$ at one loop with the appropriate value of
$\Lambda_{\rm QCD}^{(4)}$ for each set.

Previously~\cite{EKV2}, we worked at LO only since the 
EHKQS sets are evolved according to the LO DGLAP+GLRMQ
equations
using a one-loop evaluation of $\alpha_s$.  
Thus these LO distributions should generally not be mixed
with NLO matrix elements and the two-loop $\alpha_s$.  However, the charm quark
total cross section is increased and the
$p_T$ distribution is broadened at NLO relative to LO~\cite{RVkfac}.  
Thus, to determine whether or not the enhancement is experimentally measurable,
we must go beyond the ratio presented in Ref.~\cite{EKV2}.  To accomplish this,
we assume that the enhancement will be the same at NLO as at LO and
employ a NLO cross section closest to the calculation of the enhancement
in Ref.~\cite{EKV2}.  

As described in Ref.~\cite{RVkfac}, the theoretical $K$ factor may be 
defined in more than one way, depending on how the LO contribution to the 
cross section is calculated.  In all cases, the ${\cal O}(\alpha_s^3)$ 
contribution to cross section is calculated using NLO PDFs and the two-loop 
evaluation of $\alpha_s$.  If the LO contribution is also calculated using 
NLO PDFs and a two-loop $\alpha_s$, this is the ``standard NLO'' cross section.
It is used in most NLO codes, both in the global analyses of the NLO PDFs and 
in evaluations of cross sections and rates \cite{RVkfac}.  The $K$ factor 
formed when taking the ratio of the ``standard NLO'' cross section to the
LO cross section with the NLO PDFs \cite{RVkfac}, $K_0^{(1)}$, 
indicates the convergence of
terms in a fixed-order calculation \cite{klmv_cc}.
On the other hand, if the LO
contribution to the total NLO cross section employs LO PDFs and the one-loop
$\alpha_s$, we 
have a cross section which we refer to here as ``alternative NLO''.
The $K$ factor calculated taking the ratio of the ``alternative NLO'' cross section to the
LO cross section with LO PDFs \cite{RVkfac}, $K_2^{(1)}$, indicates the
convergence of the hadronic cross section toward a result.  If $K_0^{(1)} >
K_2^{(1)}$, convergence of the hadronic cross section is more likely 
\cite{klmv_cc}.  This is indeed the case for charm production \cite{RVkfac}.
We also note that $K_2^{(1)}$ is a much weaker function of energy than
$K_0^{(1)}$.
Since, in the absence of nonlinear NLO PDFs,
the ``alternative NLO'' cross section is more consistent with 
the enhancement calculated in Ref.~\cite{EKV2}, we use this cross section to
calculate the NLO $D$ meson rates and $p_T$ spectra.  We note also that,
in both cases, the $p_T$ distributions have the same slope even though
$K_2^{(1)}$, for the alternative NLO cross section, is somewhat smaller.  
Thus, using a non-standard NLO calculation will not change the slope of the
$p_T$ distributions, distorting the result.

The LO and NLO calculations used to obtain the full NLO result in 
both cases can be defined by modification of Eq.~(\ref{sigcc}).  For
simplicity, we drop the dependence of the cross section on 
$\sqrt{s}$, $m_c$, $Q_F^2$ and $Q_R^2$
on the left-hand side of Eq.~(\ref{sigcc}) in the following.
We thus define the full LO charm production cross section as
\begin{eqnarray}
 d\sigma_{\rm LO}^{\rm 1L} =
   \sum_{i,j=q,\overline q,g} 
   f_i^{\rm LO}(x_1,Q_F^2)\otimes f_j^{\rm LO}(x_2,Q_F^2)
   \otimes d\hat \sigma^{\rm LO}_{ij\rightarrow c \overline c}
(\alpha_s^{\rm 1L}(Q_R^2),x_1,x_2)
\label{sigflo}
\end{eqnarray}
where the superscript ``LO'' on $d\hat \sigma_{ij\rightarrow c \overline c}$
indicates the use of the 
LO matrix elements while the superscript ``1L'' indicates that the
one-loop expression of $\alpha_s$ is used.  The LO cross section typically used
in NLO codes employs the NLO PDFs and the two-loop (2L) $\alpha_s$ so that
\begin{eqnarray}
 d\sigma_{\rm LO}^{\rm 2L} =
   \sum_{i,j=q,\overline q,g} 
   f_i^{\rm NLO}(x_1,Q_F^2)\otimes f_j^{\rm NLO}(x_2,Q_F^2)
   \otimes d\hat \sigma^{\rm LO}_{ij\rightarrow c \overline c}
(\alpha_s^{\rm 2L}(Q_R^2),x_1,x_2) \, \, .
\label{signlolo}
\end{eqnarray}
In either case, the NLO contribution, ${\cal O}(\alpha_s^3)$ for heavy
quark production, is
\begin{eqnarray}
 d\sigma_{{\cal O}(\alpha_s^3)} =
   \sum_{i,j=q,\overline q,g} 
   f_i^{\rm NLO}(x_1,Q_F^2)\otimes f_j^{\rm NLO}(x_2,Q_F^2)
   \otimes \hspace{-0.3cm}\sum_{k=0,q,\overline q,g} d\hat 
\sigma^{\rm NLO}_{ij\rightarrow c \overline c k}(\alpha_s^{\rm 
2L}(Q_R^2),Q_F^2,x_1,x_2)
\label{signlocont}
\end{eqnarray}
where the superscript ``NLO'' on $d\hat \sigma_{ij\rightarrow c \overline c k}$
indicates the use of the NLO matrix elements.  The additional sum over $k$ in
Eq.~(\ref{signlocont}) includes the virtual, $k=0$, and real, $k = q$,
$\overline q$ or $g$ depending on $i$ and $j$, NLO corrections.
In the calculations of $d\sigma_{\rm LO}^{\rm 2L}$ and 
$d\sigma_{{\cal O}(\alpha_s^3)}$, we use the value of $\Lambda_{\rm QCD}^{(4)}$
given for the NLO PDFs and work in the ${\overline {\rm MS}}$ scheme.
The standard NLO cross section is then
\begin{eqnarray}
 d\sigma_{\rm NLO}^{\rm std} =  d\sigma_{\rm LO}^{\rm 2L} +  
d\sigma_{{\cal O}(\alpha_s^3)} \, \, 
\label{nlostd}
\end{eqnarray}
while our ``alternative NLO'' cross section is defined as
\begin{eqnarray}
 d\sigma_{\rm NLO}^{\rm alt} =  d\sigma_{\rm LO}^{\rm 1L} +  
d\sigma_{{\cal O}(\alpha_s^3)} \, \, .
\label{nloalt}
\end{eqnarray}
Since the enhancement in Ref.~\cite{EKV2} was defined using $d\sigma_{\rm
LO}^{\rm 1L}$ only, the best we can do is to  
use the alternative NLO cross section in our analysis, as described below.

We now discuss how the enhancement is taken into account in the context of the 
NLO computation.
We calculate the LO inclusive charm $p_T$ distribution, 
$d^{2}\sigma/dp_Tdy$, with the detected charm (anticharm) quark in the 
rapidity interval $\Delta y$ with $|y|<1$,
motivated by 
the  pseudorapidity acceptance of the ALICE tracking barrel, $|\eta|<0.9$.
The rapidity, $y_2$, of the undetected anticharm (charm) quark is integrated 
over. The charm enhancement factor 
$R(p_T,\Delta y)$ is then 
\begin{eqnarray}
 R(p_T,\Delta y) = \frac
{ \displaystyle \int_{\Delta y} dy \int dy_2 
 \frac{d^3 \sigma({\scriptstyle \rm EHKQS})}{dp_T dy dy_2}}
{\displaystyle \int_{\Delta y} dy\int dy_2  
 \frac{d^3 \sigma({\scriptstyle \rm CTEQ61L})}{dp_T dy dy_2}} \, \, .
\label{rofpt}
\end{eqnarray}
Numerically, this ratio is very close to
$R(p_T,y,y_2)$, computed in Ref.~\cite{EKV2}, as seen by a comparison of
$R(p_T,\Delta y)$ in Fig.~\ref{fig1} with Fig.~2 of Ref.~\cite{EKV2}.

Next, we assume that the enhancement calculated at LO is the
same when calculated at NLO.  This is a rather strong assumption but,
until the nonlinear evolution has been completely analyzed to NLO, it
is the only reasonable assumption we can make to test whether the enhancement
can be detected with ALICE which will measure the physical $\ptD$ distribution.
The alternative
NLO cross section is therefore the closest in spirit to the LO
computation in Ref.~\cite{EKV2}. Thus, the enhanced NLO charm 
$p_T$ distribution is
\begin{eqnarray}
R(p_T,\Delta y) \, \,d\sigma_{\rm NLO}^{\rm alt}(\Delta y)/dp_T \, \, .
\label{rnloalt}
\end{eqnarray}  

In our calculations, we use values of the charm quark mass and scale
that have been fit to the total cross section data using standard NLO
calculations.  The best agreement with the total cross section data is
obtained with $m_c = 1.2$~GeV and $Q^2 = 4m_c^2$ for 
DGLAP-evolved NLO PDFs such as CTEQ6M~\cite{cteq61} and MRST
\cite{mrst}.  Nearly equivalent agreement may be obtained with $m_c =
1.3$~GeV and $Q^2 = m_c^2$~\cite{HPC,rvww02}.    
Agreement with the fixed-target total
cross sections can only be achieved with higher $m_c$ by
making the factorization scale, $Q_F^2$, larger than the renormalization
scale, $Q_R^2$.  Using a lower value of $Q_R^2$ increases the cross section by
inflating $\alpha_s$.  If $Q_F^2 \leq Q_0^2$,
the PDFs are unconstrained in $Q^2$ and are thus unreliable.  
We keep $Q_F^2 = Q_R^2$ since all typical PDFs are fit using this assumption.
Thus we limit
ourselves to relatively small values of $m_c$ to obtain agreement with
the total cross section data.  

We note that while  $m_c$ is the only relevant scale in the 
total cross section, $m_T$ is used instead of 
$m_c$ in the calculations of $R$ and $d\sigma_{\rm NLO}^{\rm alt}(\Delta
y)/dp_T$ to control $p_T$-dependent logarithms at NLO
\cite{EKV2}.  Our main results
are then based on the inputs that give the best agreement with the total cross
section data, $m_c = 1.2$~GeV and $Q^2 = 4m_T^2$ as well as $m_c =
1.3$~GeV and $Q^2 = m_T^2$.  These two choices will form the baseline results
against which other parameter choices will be compared to see if the
enhancement can be detected.

\section{From charm to $D$ enhancement}

Previously~\cite{EKV2}, we did not include parton intrinsic transverse
momentum, $k_T$, broadening or fragmentation.  
Since the effect of intrinsic $k_T$ is quite small at
LHC energies, on the order of 10\% or less~\cite{HPC}, we have not
included intrinsic $k_T$ in our calculations.
To make a more realistic $D$ meson distribution, we have
modified the charm $p_T$ distribution by the heavy quark
string fragmentation in PYTHIA~\cite{pythia},
as explained below.
The resulting $D$ distribution is significantly harder
than that obtained using the Peterson fragmentation function~\cite{pete}.

We first show how the $p_T$-dependent enhancement, calculated for the
charm quark, is reflected in the $D$ meson $p_T$ distribution.  Charm
events in $pp$ collisions at $\sqrt{s} = 14$~TeV are generated using
PYTHIA (default settings) with the requirement that one of the quarks
is in the interval $|y|<1$.  The charm quarks are hadronized using the
default string model.  Since $c$ and $\overline c$ quarks fragment to
$D$ and $\overline D$ mesons\footnote{Here $D \equiv D^+, D^0$.},
respectively, in each event related $(c,D)$ and
$(\overline{c},\overline{D})$ pairs can easily be
identified\footnote{Events containing charm baryons were rejected.}.
These pairs are reweighted to match an arbitrary NLO charm quark $p_T$
distribution, $dN^c_{\rm NLO}/dp_T$.  If $dN^c_{\rm PYTHIA}/dp_T$ is
the charm $p_T$ distribution given by PYTHIA, each $(c,D)$ pair is
assigned the weight
\begin{equation}
\mathcal{W}(p_T) = \frac{dN^c_{\rm NLO}/dp_T}
                        {dN^c_{\rm PYTHIA}/dp_T} \, \, 
\end{equation}
where $p_T$ is the transverse momentum of the charm quark of the pair. 
Therefore, the reweighted final-state $D$ distribution
corresponds to the one that would be obtained by applying string fragmentation 
to the NLO $c$-quark distribution.

\begin{figure}[t]
\centering\includegraphics[width=15cm]{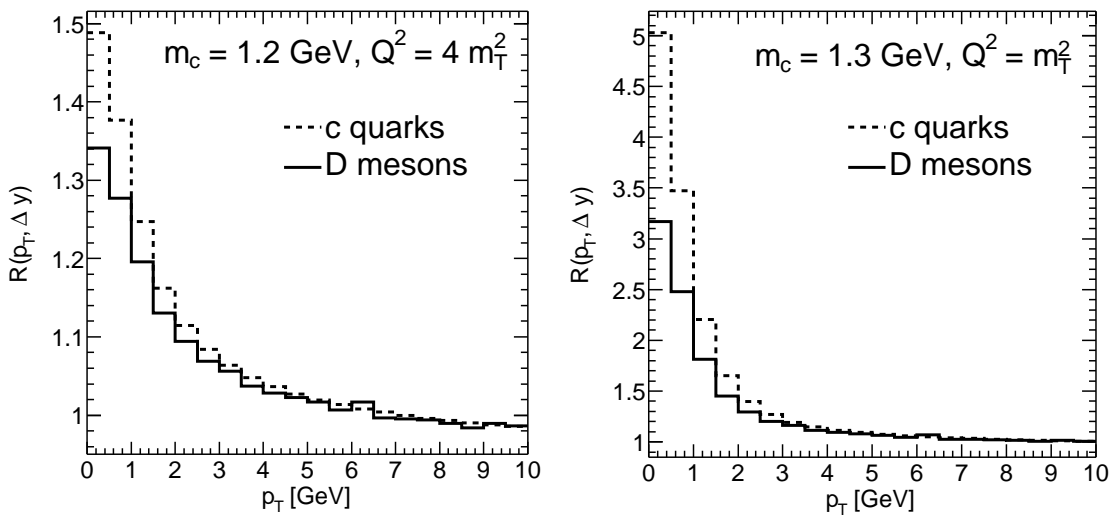}
\caption[]{\small Enhancement factor $R(p_T,\Delta y)$ for charm quarks 
(dashed histogram) and for \mbox{$D~(\equiv D^+,D^0)$} 
mesons (solid histogram), obtained
after PYTHIA string fragmentation.  The left-hand side shows the result for 
$m_c = 1.2$~GeV and
$Q^2 = 4m_T^2$ while the right-hand side is the result for $m_c = 1.3$~GeV and
$Q^2 = m_T^2$.
}
\label{fig1}
\end{figure}

In Fig.~\ref{fig1} we compare the enhancement factor $R$, calculated in
Eq.~(\ref{rofpt}) for $c$ quarks and $D$ mesons generated from the weighted
PYTHIA charm distributions. The two cases described previously, 
$m_c=1.2$~GeV, $Q^2=4m_T^2$ (left-hand side) and $m_c=1.3$~GeV, $Q^2=m_T^2$ 
(right-hand side) are considered. In both cases, the enhancement survives 
after fragmentation. It is interesting to note that the $D$ 
enhancement is somewhat lower than that of the charm: in the most optimistic 
case, the factor of five charm enhancement has reduced to a factor of three for
the $D$ mesons. This occurs because, for a given
$\ptD$, the $D$ spectrum receives contributions from charm quarks with
$p_T \, \gsim \, \ptD$, where the charm enhancement is smaller. 
The $D$ enhancement also vanishes with increasing transverse momenta, like
the charm enhancement.

\section{$D^0$ reconstruction in $pp$ collisions with ALICE}

The transverse momentum distribution of $D^0$ mesons produced at central 
rapidity, $|y|<1$, can be directly measured from the exclusive 
reconstruction of $D^0\to K^-\pi^+$ decays (and charge conjugates) 
in the Inner Tracking System (ITS), Time Projection Chamber (TPC) and 
Time Of Flight (TOF) detectors of the ALICE barrel, 
$|\eta|<0.9$~\cite{aliceTP}. 
The main feature of the $D^0$ decay topology is the presence of two tracks
displaced 
from the interaction point by, on average, 50~$\mu$m, for 
$\ptD\simeq 0.5$~GeV, to 120~$\mu$m, for $\ptD\gsim 5$~GeV.
Such displacement can be resolved with the ALICE tracking detectors
and thus a
large fraction of the combinatorial background in the $K^\mp \pi^\pm$ invariant
mass distribution can be rejected.
The low value of the magnetic field, 0.4~T, and the 
$K/\pi$ separation in the TOF detector extend
the $D^0$ measurement down to $\ptD \sim 0$. The analysis strategy 
and the pertinent selection cuts were studied with a realistic, detailed 
simulation of the detector geometry and response, including the main
background sources~\cite{thesisAD,D0jpg}. 

\begin{figure}[!t]
\centering\includegraphics[width=0.7\textwidth]{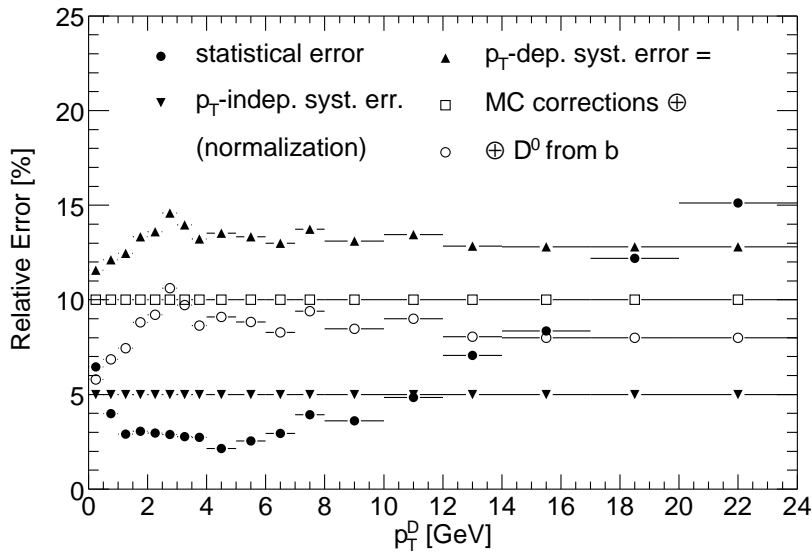}
\caption[]{\small Estimated relative uncertainties on the measurement of the 
$D^0$ differential cross section in $pp$ collisions at the LHC 
with ALICE~\cite{thesisAD}. 
Statistical uncertainties correspond to $10^9$ minimum-bias 
$pp$ events (an $\approx 9$ month run with a luminosity of $\approx 5 \times
10^{30}~{\rm cm}^{-2}{\rm s}^{-1}$).
}
\label{fig2}
\end{figure}

The expected ALICE performance for $pp$ collisions at $\sqrt{s}=14$~TeV
is summarized in Fig.~\ref{fig2} where the estimated relative uncertainties 
are reported as a function of $\ptD$.  The main contributions to the 
$p_T$-dependent systematic error (triangles)
are the detector acceptance and reconstruction
efficiency corrections (squares), $\simeq 10\%$,
and the correction for feed-down from 
bottom decays, $B\rightarrow D^0+X$ (open circles), $\simeq 8\%$. 
The latter is estimated based on the present $70-80$\%
theoretical uncertainty in the $b \overline b$ 
cross section at LHC energies~\cite{yrepHVQ}. However, we expect this 
uncertainty to be significantly reduced by the measurement of $B$ decays to
single electrons, $B\to e^\pm+X$, in ALICE~\cite{yrepHVQ}. 
The $p_T$-independent systematic 
error introduced by normalization to the 
$pp$ inelastic cross section (inverted triangles)
is also reported.  This cross section will be 
measured by the TOTEM experiment~\cite{totem} with an $\simeq 5\%$ uncertainty.

The statistical error corresponding to $10^9$ minimum-bias $pp$ events (filled
circles), an $\approx 9$ month run with a luminosity of $\approx 5 \times
10^{30}~{\rm cm}^{-2}{\rm s}^{-1}$, is smaller than or on the order of the 
$p_T$-dependent systematic error up to $\ptD\simeq 24~{\rm GeV}$ for the 
alternative NLO cross section calculated using $m_c=1.2$~GeV, $Q^2=4m_T^2$ and 
the CTEQ6 PDFs with no enhancement.  The relative 
statistical error depends on the charm cross section, as we now explain.  
For a given $D^0$ $\ptD$ or $\ptD$ range, the statistical error is the error on
the number of real $D^0$ ($\overline D^0$) mesons in the $K^\mp \pi^\pm$
invariant mass distribution, the signal, ${\rm S}(\ptD)$. 
The error is equal to
$\sqrt{{\rm S}(\ptD) + {\rm B}(\ptD)}/{\rm S}(\ptD)$ where ${\rm B}(\ptD)$
is the
number of background candidates in the $D^0$ mass region.  Then, at low $\ptD$,
the error is $\approx \sqrt{{\rm B}(\ptD)}/{\rm S}(\ptD) \propto 
1/(d\sigma_{\scriptscriptstyle D}/d\ptD)$ 
since the invariant mass distribution is dominated by
combinatorial background.  At high $\ptD$, the background is negligible and
the error becomes $\approx 1/\sqrt{{\rm S}(\ptD)} \propto
1/\sqrt{d\sigma_{\scriptscriptstyle D}/d\ptD}$. 
In our subsequent results, the statistical errors are 
calculated taking this cross section dependence into account.

\section{Sensitivity to the enhancement}

\begin{figure}[!t]
\centering\includegraphics[width=15cm]{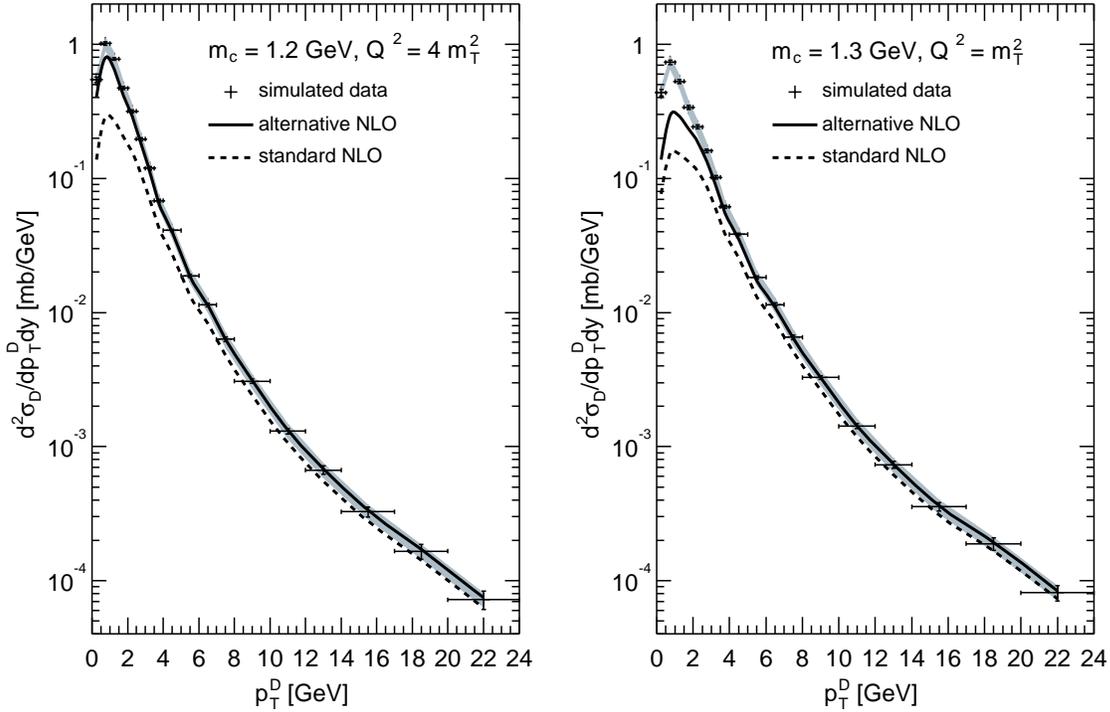}
\caption[]{\small Comparison of the simulated ALICE data generated from
$R(p_T,\Delta y)
d\sigma_{\rm NLO}^{\rm alt}$ with the alternative (solid) and standard (dashed)
NLO calculations.  The effect of string fragmentation is included in the 
``data'' points as well as in the curves.
The left-hand side shows the result for $m_c = 1.2$~GeV and
$Q^2 = 4m_T^2$ while the right-hand side is the result for $m_c = 1.3$~GeV and
$Q^2 = m_T^2$. 
The error bars on the data represent the statistical 
error and the shaded band represents the $p_T$-dependent systematic error.
The 5\% normalization error is not shown.
}
\label{fig3}
\end{figure}

Figure~\ref{fig3} shows the double-differential $D^0$ cross section, 
$d^2\sigma_{\scriptscriptstyle D}/d\ptD dy$, 
in $|y|<1$ as a function of the transverse momentum.
The points represent
the expected ``data'' measured by ALICE, obtained from the 
alternative NLO cross section scaled by the enhancement factor 
$R(p_T,\Delta y)$ defined in Eq.~(\ref{rofpt}), and modified by string 
fragmentation. The solid and dashed curves are obtained by applying string
fragmentation to the alternative NLO and standard 
NLO $c\overline c$ cross sections, respectively. 
Thus, the ``data'' points include the enhancement while the curves do not.
The horizontal error bars indicate the bin 
width, the vertical error bars represent the statistical error and the shaded 
band gives the $p_T$-dependent systematic error. The 5\% $p_T$-independent
systematic error on 
the normalization is not shown.  The left-hand side shows the results for
$m_c = 1.2$~GeV and $Q^2 = 4m_T^2$ while the right-hand side shows those for
$m_c = 1.3$~GeV and $Q^2 = m_T^2$. 
The standard NLO cross section, Eq.~(\ref{nlostd}), and the 
$\mathcal{O}(\alpha_s^3)$ contribution to the alternative
NLO cross section, Eq.~(\ref{signlocont}), were calculated 
using the HVQMNR code~\cite{MNR} with 
CTEQ6M and $\Lambda_{\rm QCD}^{(4)}
= 0.326$ GeV. The LO contribution to the 
alternative NLO cross section, Eq.~(\ref{sigflo}), was calculated using the
CTEQ61L PDFs.  Fragmentation was included as described in Section~3.
The enhancement, the difference between the data and the solid 
curve visible for $\ptD \, \lsim \, 3$~GeV, is more
pronounced for the larger mass and lower scale, shown on the right-hand side
of Fig.~\ref{fig3}.

There is a significant difference between the alternative and standard NLO 
distributions.  Part of the difference is due to the one- and two-loop
evaluations of $\alpha_s$ since $\alpha_s^{\rm 2L} < \alpha_s^{\rm 1L}$.
This decrease will in turn reduce the ${\cal O}(\alpha_s^3)$ 
contribution to the alternative NLO
result relative to the LO component of Eq.~(\ref{sigflo}).  
In addition, the standard NLO
cross section would be reduced overall relative to a calculation with the
same $\Lambda_{\rm QCD}^{(4)}$ at LO and NLO.  However, these factors alone
cannot explain the rather large difference between the standard and
alternative NLO cross sections at low $\ptD$.  The most important contribution
is the large differences between the LO and NLO gluon distributions, 
especially at low
scales. The slope of the CTEQ61L gluon distribution at $Q^2 = 1.69$~GeV$^2$
with $x$ is very small until $x > 0.01$.  On the other hand, the CTEQ6M gluon
$x$ slope is large and has the opposite sign relative to CTEQ61L 
for $x < 0.04$.  The ratio of the two sets at $x \approx
10^{-5}$ is very large, CTEQ61L/CTEQ6M $\approx 100$.  At $Q^2 = 5.76$~GeV$^2$,
the scale corresponding to $4m_c^2$ with $m_c = 1.2$~GeV, this ratio decreases
to a factor of two.  We note that at fixed-target energies, $\sqrt{s} \leq
40$~GeV, the standard and 
alternative NLO results are indistinguishable from each other since the 
LO and NLO gluon
distributions are rather similar in this relatively high $x$ region, $0.05 \leq
x \leq 0.1$.

In order to address the question of the experimental sensitivity to the
effect of nonlinear gluon evolution on low-$p_T$ charm production, we 
consider, as a function of $\ptD$, 
the ratio of the simulated data, including the enhancement, 
to alternative NLO calculations using a range of $m_c$ and $Q^2$
along with PYTHIA string fragmentation.  We denote this ratio as 
``Data/Theory''. Thus, given the 
measured $D^0$ $p_T$ distribution,
we try to reproduce this result with NLO calculations employing
recent linearly-evolved PDFs and tuning $m_c$ and 
$Q^2$.  We note that these parameters are not really free but are bounded by
the range $1.2\, \lsim \, m_c \, \lsim \, 1.8$~GeV and $1\, \lsim \, 
Q^2/m_T^2\, \lsim \, 4$, as
described in Section 2 and in Ref.~\cite{EKV2}. 

Since the enhancement has disappeared
for $\ptD \, \gsim \, 5$~GeV, 
we refer to this unenhanced region as high $\ptD$.
The $\ptD$ region below 5~GeV, 
where the enhancement is important, is referred 
to as low $\ptD$.  If no set of parameters can
describe both the high- and low-$\ptD$ components
of the distribution equally 
well, and, if 
the set that best reproduces the high-$\ptD$ part underestimates the low-$\ptD$
part, this would be a strong indication of the presence of nonlinear effects.

\begin{figure}[!t]
\centering\includegraphics[width=\textwidth]{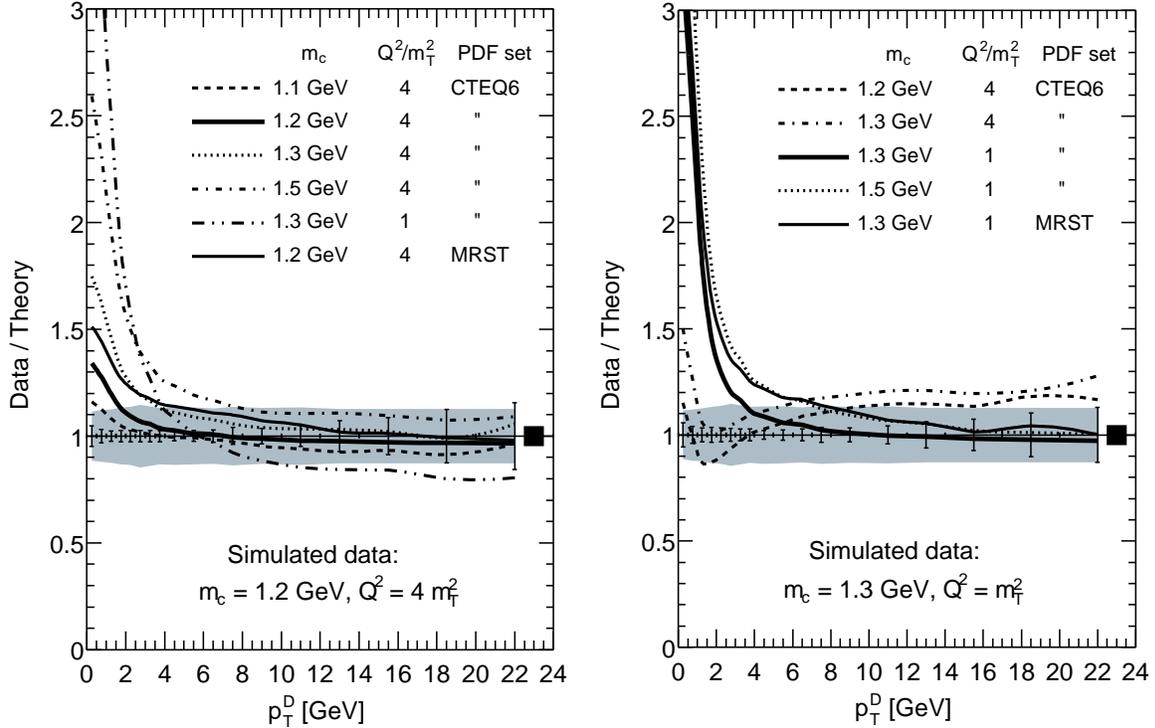}
\caption[]{\small Ratio of the generated ALICE data relative to
calculations of the alternative NLO cross sections with several sets of
parameters 
and PYTHIA string fragmentation.
The left-hand side shows the result for $m_c = 1.2$~GeV and
$Q^2 = 4m_T^2$ while the right-hand side is the result for $m_c = 1.3$~GeV and
$Q^2 = m_T^2$. 
}
\label{fig4}
\end{figure}

The Data/Theory plots are shown in Fig.~\ref{fig4}.
The points with the statistical (vertical bars) and $p_T$-dependent systematic 
(shaded region) error correspond to the data of Fig.~\ref{fig3}, 
including the enhancement, divided by themselves, depicting the sensitivity
to the theory calculations. The black squares on 
the right-hand sides of the lines ${\rm Data/Theory=1}$ 
represent the 5\% $p_T$-independent error 
on the ratio coming from the cross section normalization. As clearly shown 
in Fig.~\ref{fig2}, this error is, 
however, negligible with respect to the present estimates of the 
other systematic uncertainties ($\simeq 13\%$).

On the left-hand side, the thick solid curve with $m_c =
1.2$~GeV and $Q^2 = 4m_T^2$ best agrees with the high-$\ptD$ ratio by
construction since $R \approx 1$ at large $\ptD$.  
It also shows the effect of the enhancement 
well beyond the error band for $\ptD\, \lsim \, 2$~GeV.
Better agreement with the data over the entire $\ptD$ 
range can be achieved only
by choosing a charm quark mass lower than 1.2~GeV, below the nominal range of
charm masses, as illustrated by the dashed curve for $m_c=1.1$~GeV.
Higher masses with $Q^2 = 4m_T^2$ produce much larger
Data/Theory ratios than the input distribution. Choosing e.g.\ $m_c = 1.8$~GeV 
(not shown) would give
a larger Data/Theory ratio than the $m_c = 1.5$~GeV result (dot-dashed curve). 
The ratio with $m_c =
1.3$~GeV and $Q^2 = m_T^2$ (dot-dot-dashed curve) gives a much larger 
ratio at low $\ptD$ and drops
below the data for $\ptD > 8$~GeV.  

We have checked how the results change when the
renormalization and factorization scales are separated.  When $m_c = 1.3$~GeV,
$Q_R^2 = m_T^2$ and $Q_F^2 = 4m_T^2$, the faster evolution of the higher
$Q_F^2$ and the larger $\alpha_s(Q_R^2)$ resulting from the lower $Q_R^2$
leads to reasonable agreement between data and theory at low $\ptD$. 
However, at
high $\ptD$, the theory distribution is harder so that the Data/Theory
ratio drops below the error band for $\ptD > 2$ GeV.  On the other hand,
when $m_c = 1.3$~GeV, $Q_R^2 = 4m_T^2$ and $Q_F^2 = m_T^2$, the theory
cross section is reduced relative to the data and the Data/Theory ratio is
above the error band over all $\ptD$.

We also present the ratio using the MRST
parton densities (MRST2001 LO~\cite{MRST2001} in Eq.~(\ref{sigflo}) and 
MRST2002 NLO~\cite{MRSTNNLO} in Eq.~(\ref{signlocont}))
with $m_c = 1.2$~GeV and $Q^2 = 4m_T^2$.  We find that this
result, the thin solid curve, 
also agrees reasonably well with the CTEQ6 results shown in the thick solid
curve for the same $m_c$ and $Q^2$.  Thus, the enhancement seems to be 
rather independent of the PDF.
The CTEQ61L and the MRST2001 LO distributions are similar at low $x$,
suggesting that PDFs based on this MRST set would produce an enhancement
like that of Ref.~\cite{EKV2}.  However, the MRST2002 NLO and CTEQ6M NLO gluon
distributions are very different at low $x$.  The MRST2002 NLO gluon 
distribution is negative at low scales while the CTEQ6M gluon distribution 
goes to zero as $x\rightarrow 0$.  
Thus the effects of nonlinear evolution at NLO could be 
considerably different.

On the right-hand side of Fig.~\ref{fig4}, with $m_c = 1.3$~GeV and $Q^2 =
m_T^2$, the thick solid curve,
employing the same parameters as the data,
gives the best agreement at high $\ptD$.  
We note that even though the results with $Q^2 = 4m_T^2$ and 
$m_c \leq 1.3$~GeV lie closer to the data at low $\ptD$ and within the 
combined statistical and systematic error at higher $\ptD$, 
the curves with these parameters have the wrong slopes for 
$\ptD\lsim 8$~GeV. 
The systematic errors are more likely to shift the data points up or down 
as a whole rather than twist the $\ptD$ shape.  The statistical sensitivity 
is expected to be good enough to distinguish the difference in curvature.
Varying $Q_F^2$ and $Q_R^2$ separately results in similarly poor
agreement as that noted for $m_c = 1.2$~GeV and $Q^2 = 4m_T^2$.
Finally, the results obtained with the MRST PDFs, shown in the thin 
solid line, 
do not alter the conclusions already drawn for CTEQ6.

\section{Conclusions}

With constraints from HERA, the nonlinear DGLAP+GLRMQ evolution
at LO leads to an enhancement of the free proton gluon distributions at 
$x \, \lsim 
\, 0.01$ and $Q^2\, \lsim \, 10$~GeV$^2$ relative to DGLAP-evolved LO sets 
such as CTEQ61L. Consequently, charm 
hadroproduction at $\sqrt{s}\, \gsim \, 1$~TeV 
should be larger than expected from
DGLAP-evolved PDFs alone~\cite{EKV2}. In this paper, we have studied whether 
the EHKQS gluon distributions~\cite{EHKQS} could generate an observable $D$
meson enhancement in $pp$ collisions at the LHC.
Since larger $x$ values are probed at lower energy colliders, the 
enhancement described here would be reduced. At RHIC, $\sqrt{s}=200$~GeV,
the effect is too small to be reliably observed. However, $D$ mesurements
at the Tevatron, $\sqrt{s}=1.96$~TeV, may allow to detect an enhancement 
if the minimum $\ptD$ was lowerd to $\approx 1$~GeV.

In order to consider more realistic $\ptD$ distributions and yields, we have 
calculated the NLO contribution to charm production 
using the HVQMNR code~\cite{MNR}. 
Since the LO EHKQS PDFs cannot be used consistently with the NLO matrix
elements, we assume the charm enhancement is the same at LO and NLO. 
We note that nonlinear effects on the NLO gluon 
distributions may be smaller than at LO, thus reducing the NLO charm
enhancement.  Therefore, our results may be considered upper limits of the NLO
$D$ enhancement. Note also that if NLO DGLAP+GLRMQ PDFs 
that fit the small-$x$ and small-$Q^2$ HERA data were available,
it would be possible to base our analysis on the standard NLO charm
cross section instead of the ``alternative NLO'' result defined in
Eq.~(\ref{nloalt}).
Improved gluon distributions at low $x$ and
$Q^2$ may make the standard and alternative NLO results more similar at high
energies, as they are at lower $\sqrt{s}$ where $x$ is larger.

Using the EHKQS LO PDFs and LO matrix elements for charm quark
production and PYTHIA string fragmentation for $D$ meson hadronization, we 
have demonstrated that more than half of the charm enhancement relative to
calculations with the CTEQ61L LO PDFs indeed survives to the $D$ mesons. In 
the most optimistic case, $m_c=1.3$ GeV and $Q^2=m_T^2$, the factor of five
charm enhancement at $|y|\le 1$ and $p_T\to 0$
is reduced to a factor of three at $\ptD\to 0$.
For larger values of $m_c$ and $Q^2$, the charm enhancement is smaller 
because the gluon enhancement due to nonlinear evolution 
decreases with increasing $Q^2$.

The $D$ meson enhancement, however, drops rapidly with transverse momentum 
so that for $\ptD\sim 5$ GeV it is only a few percent. Therefore, $D$
measurement capability at small $\ptD$ is necessary to verify the effect 
experimentally.  The ALICE detector can do this through direct $D^0$ 
reconstruction in the $K^-\pi^+$ decay channel. We have demonstrated, using the
error analysis of Ref.~\cite{thesisAD}, that, in the most optimistic 
case, the enhancement can be detected above the experimental statistical and 
systematic errors.  The sensitivity of the $D$ enhancement to the scale 
has also been considered and we have shown that when the charm 
mass is somewhat smaller,
$m_c=1.2$ GeV, but the scale is larger, $Q^2=4m_T^2$, it is more difficult to
detect the enhancement over the experimental uncertainties. 
The ALICE sensitivity to $D$ meson production at very low transverse momentum 
may further improve by combining the $D^0\to K^-\pi^+$ measurement with   
those of $D^+ \to K^- \pi^+ \pi^+$ and $D^0\to K^-\pi^+\rho^0$. 
A fast-simulation feasibility study of $D^+ \to K^- \pi^+ \pi^+$ 
reconstruction~\cite{aliceitstdr} indicates that a
performance similar to that of $D^0\to K^-\pi^+$ could be achieved. More 
detailed analyses, currently in progress, will assess the low-$\ptD$ reach 
of this channel. 
 
\bigskip\bigskip
\noindent {\bf Acknowledgments:}
The work of A.D. and M.B. was carried out within the ALICE Collaboration,
of which they are members, and using the software framework developed by 
the off-line project. A.D. and M.B. acknowledge the ALICE off-line
group and  the Physics Coordinator ${\rm K.~\check{S}afa\check{r}}$\'ik
for support and useful discussions.
The work of R.V. was supported in part by the Director, Office of
Energy Research, Division of Nuclear Physics of the Office of High
Energy and Nuclear Physics of the U. S.  Department of Energy under
Contract Number DE-AC03-76SF00098.  
K.J.E. and V.J.K. gratefully acknowledge the financial support 
from the Academy of Finland, projects 50338, 80385 and 206024.

\end{document}